\documentstyle[aps,prb]{revtex} 
\input epsf.sty

\begin{document}
\setcounter{figure}{0} 
\renewcommand{\theequation}{\arabic{equation}} 

\twocolumn[ 
\hsize\textwidth\columnwidth\hsize\csname@twocolumnfalse\endcsname 
 
\draft 

\title{N\'eel-Dimer Transition in Antiferromagnetic Heisenberg Model \\
 and
Deconfinement of Spinons at the Critical Point} 
\author{Daisuke Yoshioka, 
Gaku Arakawa and Ikuo Ichinose} 
\address{Department of Applied Physics, Graduate School of Engineering, \\
Nagoya Institute of Technology, 
Nagoya, 466-8555 Japan 
} 
\author{Tetsuo Matsui} 
\address{Department of Physics, Kinki University, 
Higashi-Osaka, 577-8502 Japan 
} 
\date{\today}  
 
\maketitle

\begin{abstract}   
Quantum phase transition from the N\'eel to the dimer states
in an antiferromagnetic(AF) Heisenberg model on 
square lattice is studied.
We introduce a control parameter $\alpha$ for the exchange coupling
which connects the N\'eel ($\alpha=0$) and the dimer ($\alpha=1$) states.
We employ the $CP^1$ representation of the 
$s={1\over 2}$ spin operator and integrate out the half of 
the $CP^1$ variables
at odd sites to obtain a $CP^1$ nonlinear $\sigma$ model.
The effective coupling constant is a function of $\alpha$ and at $\alpha=0$
the $CP^1$ model is in the ordered phase which corresponds to the N\'eel
state of the AF Heisenberg model.
A phase transition to the dimer state occurs at a certain critical
value of $\alpha_C$ as $\alpha$ increases.
In the N\'eel state, the dynamical composite U(1) gauge field in the $CP^1$
model is in a Higgs phase and low-energy excitations are gapless spin
wave.
In the dimer phase, a confinement phase of the
gauge theory with $s=1$ excitations is realized.
For the critical point, we argue that a deconfinement phase, which is 
similar to the Coulomb phase in $3$ spatial dimensions,
is realized and $s={1\over 2}$ spinons
appear as low-energy excitations.

\end{abstract} 
 

]%the end of twocolumn 

%%%%%%%%%%%%%%%%%%%%%%%%%%%%%%%%%%%%%%%%%%%%%%%%%%%%%%%%%%% 
 
\setcounter{footnote}{0}
\setcounter{equation}{0} 
%%%%%%%%%%%%%%%%%%%%%%%%%%%%%%%%%%%%%%%%%%%%%%%%%%%%%%%%%%% 
%\onecolumn
%\Large
%\section{Introduction}
Quantum phase transition(QPT) is one of the most interesting problem in
these days.\cite{QPT}
It is often argued that the simple Ginzburg-Landau theory does not
apply to certain class of the QPT's. 
In this paper we shall study $s={1\over 2}$ antiferromagnetic(AF)
Heisenberg model on $2$-dimensional square lattice with nonuniform 
exchange couplings,
\begin{equation}
H_{AF}=\sum_{x,j}J_{xj}\vec{S}_x\cdot\vec{S}_{x+j}
\label{AFH}
\end{equation}
where $x$ denotes site of the spatial lattice, $j$ is the direction
index ($j=1,2$) and $\vec{S}_x$ is the spin operator at site $x$.
We rename the even lattice sites $x=(o,i)$ where $o$ denotes odd site
and the index $i=1,2,3$ and $4$ specifies its four nearest-neighbor(NN)
even sites (see Fig.1).

%---------------------------------------------------
% 2004.3.24 insert fig-1. By daisuke
\begin{figure}[htbp]
\begin{center}
\leavevmode
\epsfxsize=6.5cm
\epsffile{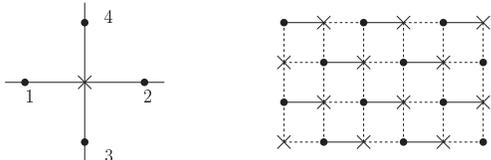}
\caption{Numbering and dimer picture:cross symbol is odd site, and spot
 symbol is even site. Solid line means that its correlation is stronger 
than dotted lines.}
\label{fig-1}
\end{center}
\end{figure}
%----------------------------------------------------

The exchange couplings $J_{xj}=J_{oi}$ are position dependent and we explicitly
consider the following case which corresponds to the dimer configuration,
\begin{eqnarray}
J_{oi}&=&J+\Delta J_{oi},   \nonumber  \\
\Delta J_{oi}&=&\left\{\begin{array}{ll}
              \Delta J_{oi}=\alpha J, & i=1 \\
              \Delta J_{oi}=-\alpha J, & i=2,3,4
                      \end{array}
    \right.
\label{J}
\end{eqnarray}
where $0\le\alpha\le 1$ is a control parameter which connects the uniform
Heisenberg model to the dimer model.

It is not so difficult to derive the $CP^1$ field-theory 
model\cite{CP,CP2} from 
Eq.(\ref{AFH})\cite{IM}.
The spin operator $\vec{S}_{x}$ can be expressed in terms of the
$CP^1$ variable $z_{x}=(z^1_x,z^2_x)^t$ as 
\begin{equation}
\vec{S}_x={1\over 2}{z}^\dagger_x\vec{\sigma}z_x,
\label{S}
\end{equation}
where $\vec{\sigma}$ are the Pauli matrices and the $CP^1$ constraint
$\sum_{a=1,2}|z^a_{x}|^2=1$ guarantees the magnitude of the localized spin 
as ${1\over 2}$.

From our assignment of $J_{oi}$ (\ref{J}), it is obvious that 
$J_{o1}$ is larger than the others.
We use the path-integral formalism and
parameterize the $CP^1$ variable $z_o$ by refering to
$z_{o1}\equiv z_e$,
\begin{equation}
z_o=p_oz_e+\sqrt{1-|p_o|^2}e^{i\theta}\tilde{z}_e,
\label{parameter}
\end{equation}
where $p_o$ is a parameter, $e^{i\theta}$ is a phase
factor and $\tilde{z}_e=i \sigma_2 z^\ast_e$.
At vanishing temperature$(T)$, spins tend to point
antiparallel their NN spins, and then the parameter
$p_o$ can be treated as a small parameter.
We expand $\sqrt{1-|p_o|^2}\simeq 1-{1\over 2}|p_o|^2+\cdots$
and retain only terms up to quadratic of $p_o$.
Then we perform the Gaussian integration of $p_o$'s to obtain an effective
model of $z_e$'s for which smooth configurations dominate at $T=0$.

Calculation is rather long but straightforward\cite{IM}
and we obtain an effective
field theory of the AF Heisenberg model under study,
\begin{equation}
{\cal L}_{CP}=\sum_{\mu=\tau,x}\Big[|D_\mu z_x|^2+
\sigma_x\Big(|z_x|^2-{1\over f_{\mbox{\footnotesize{eff}}}}\Big)\Big]
+{\cal L_B},
\label{CP}
\end{equation}
where $D_\mu z_x=(\partial_\mu+iA_\mu)z_x$, $A_\mu=i{z}^\dagger\partial_\mu z$ 
is the U(1) gauge field,
$\sigma_x$ is the Lagrange multiplier for the $CP^1$ constraint
and ${\cal L_B}$ is the Berry phase term.
We have rescaled 
the imaginary time $\tau$ and $x$ as taking the continuum 
limit(the spatial coordinate $x$ in Eq.(\ref{CP}) denotes the {\em even site}
of the {\em original} lattice).
The effective coupling constant $f_{\mbox{\footnotesize{eff}}}$ and 
the ``speed of light" $c$, which is often set unity, are explicitly
given as 
\begin{eqnarray}
{1\over f_{\mbox{\footnotesize{eff}}}}&=&{1\over 2\sqrt{2}a}\cdot
{1-\alpha\over 2-\alpha}\sqrt{2(2+\alpha)\over 1-\alpha}
\equiv {1\over f_{\mbox{\footnotesize{AFH}}}}{1\over b(\alpha)},
\label{fb} \\
c&=&{\sqrt{2}aJ \over \hbar}\cdot\sqrt{{(2-\alpha)(1-\alpha) \over 2}}
\equiv c_{\mbox{\footnotesize{AFH}}} \gamma(\alpha),
\label{c}
\end{eqnarray}
where $a$ is the lattice spacing of the original lattice
and $f_{\mbox{\footnotesize{AFH}}}=2\sqrt{2}a$ and
$c_{\mbox{\footnotesize{AFH}}}={\sqrt{2}aJ \over \hbar}$.

Here we should comment on the Berry phase terms, ${\cal L_B}$,
which appear in the $CP^1$
nonlinear-$\sigma$ model representation of the AF Heisenberg model.
First we consider AF spin chains in one dimension in order to obtain 
important insight for the effect of the Berry phase terms.
For the uniform AF chains, the Berry phase reduces to the $\theta$-term
$(i\theta\int d\tau dx F_{\tau x})$ with $\theta=\pi$.
The gauge dynamics crucially depends on the value of $\theta$ and only the
case of $\theta=\pi\;(\mbox{mod}\; 2\pi)$, the deconfinement phase is realized.
Topologically nontrivial configurations are suppressed by the 
$\theta$-term when $\theta=\pi$.

On the other hand for the bond-alternating(BA) AF Heisenberg chain with
$J_x=J+(-1)^x\Delta J$, the value of the $\theta$ parameter is
$\theta={J-\Delta J \over J+\Delta J}\pi$, and therefore the above
suppression by the $\theta$-term does not occur and gauge dynamics
is essentially the same with the $\theta=0$ case.
Then the confinement of the spinon $z^a_x$ occurs, and 
low-energy excitations are gapful $s=1$ excitations 
$\vec{n}_x={1\over 2}{z}^\dagger_x\vec{\sigma}z_x$ even for 
a very small value of $\Delta J$.
This is nothing but the spin-Peierls transition.

For the $2$-dimensional uniform AF Heisenberg model, the Berry phase term
similarly appears as ${\cal L_B}={i\over 2}\sum_x \epsilon_x {\cal A}_x$,
where $\epsilon_x=1(-1)$ for even (odd) site $x$ (of the lattice with the
lattice spacing $\sqrt{2}a$ for the $CP^1$ model (\ref{CP})) and ${\cal A}_x$
is the area enclosed by the curve given by the time evolution of 
$\vec{n}_x(\tau)$.
It can be proved that it gives only vanishing contribution 
for an arbitrary smooth configuration.
However for singular configurations like instantons (monopoles),
the Berry phase is nonvanishing and it gives terms like 
$
{\cal L_B}=i{\pi \over 2}\sum_{i=1}^4\sum_x \zeta^i_x m^i_x,
$
where $\zeta^i_x$ are $0,1,2,3$ $(i=1,2,3,4)$, respectively and 
$m^i_x$ is the instanton number.\cite{Berry,Berry1} 
(The index $i$ refers to the four dual
sublattices of the original lattice.)
From the above ${\cal L_B}$, it is obvious that the cancellation mechanism of 
the instanton contribution occurs by the destructive interference
unless $m^i_x=4$ $(\mbox{mod} \;4)$.

For the nonuniform cases which are considered in this paper, 
the coefficient of the Berry phase term becomes irrational as in the case of 
the BA spin
chains and therefore cancellation of instanton effect does not occur
or at least weakens.
Then we can expect that the (global) phase structure of the 
effective gauge model
for the nonuniform AF Heisenberg model is the same with that of the
$CP^1$ model without the Berry phase terms.
More comments will be given after showing results of nonperturbative studies 
on the gauge dynamics.

It is straightforward to obtain the effective potential of 
$\sigma=\langle\sigma_x\rangle$ and 
$z^2=\langle z^2_x\rangle$ by integrating out $z^1_x$,
$$
V_{\mbox{\footnotesize{eff}}}=\sigma\Big(|z^2|^2
-{1\over f_{\mbox{\footnotesize{eff}}}}\Big)  %\nonumber  \\
+{1\over 6\pi}\Big[(\sigma+\Lambda^2)\sqrt{\sigma+\Lambda^2}-\Lambda^3
-\sigma\sqrt{\sigma}\Big],
$$
%\label{V}
%\end{equation}
where the cutoff $\Lambda={\sqrt{2\pi} \over a}$.
The effective potential $V_{\mbox{\footnotesize{eff}}}$ 
indicates that there exists a critical
coupling $f_C={4\pi \over \Lambda}$.
The existence of the phase transition has been verified by the numerical
calculation of the equivalent $O(3)$ nonlinear $\sigma$ model in
$(2+1)$ dimensions\cite{O3}.
In the weak-coupling region
$f_{\mbox{\footnotesize{eff}}}<f_C$, the spontaneous symmetry breaking
occurs and $\langle z^2_x \rangle \neq 0$.
As a result, the Higgs phase is realized in the gauge-theory terminology.
Low-energy excitations are gapless spin waves which are described by $z^1_x$.
In the strong-coupling phase $f_{\mbox{\footnotesize{eff}}}>f_C$,
$\langle \sigma_x \rangle \neq 0$ whereas $\langle z^2_x \rangle = 0$.
Local Maxwell terms appear in the effective action of the gauge field $A_\mu$,
and the confinement phase is realized.
Low-energy excitations are $s=1$ composite of the spinons which correspond to 
$\vec{n}_x$ with (mass)$^2$ 
$\propto \langle \sigma_x \rangle$.

The effective coupling $f_{\mbox{\footnotesize{eff}}}$ in (\ref{fb})
first decreases as $\alpha$ increases but above certain value of $\alpha$
it starts to increase and goes to infinity at the dimer limit $\alpha=1$.
In the uniform case $\alpha=0$, 
$f_{\mbox{\footnotesize{eff}}}(\alpha=0)<f_C$ and this means that 
the ordered N\'eel state is realized at the vanishing $T$
in the AF Heisenberg model in two spatial dimensions as
it is now widely believed.
The behavior of the effective coupling $f_{\mbox{\footnotesize{eff}}}(\alpha)$
shows the existence of a critical value $\alpha_C$ at which 
the phase transition occurs.
This result indicates that the strong-coupling phase of the $CP^1$
model (\ref{CP}) corresponds to the dimer phase in which the ground
state is nothing but spin-singlet pairs formed by the alternative strong
exchange couplings and excitations have $s=1$. 
In fact this N\'eel-dimer transition was observed by the numerical 
calculations some years ago\cite{richter}.

Hereafter we are interested in the critical point at 
$f_{\mbox{\footnotesize{eff}}}=f_C$ which separates
the N\'eel and dimer phases.
In order to investigate that ``phase", study on the gauge dynamics is
required.
At $f_C$, $\langle \sigma_x\rangle=\langle z^a_x \rangle=0$.
Effective action of the gauge field and the field $\sigma_x$
is obtained by integrating out the gapless spinon field $z^a_x\; (a=1,2)$.
The resultant effective action becomes nonlocal and therefore it is possible
for the gauge dynamics to belong to {\em different} universality class
from that of the usual gauge theory in $2+1$ dimensions.
By the continuum field-theory calculation, the effective action of the
gauge field $A_\mu$ is obtained as
\begin{equation}
{\cal L}_A\propto\int d^3x \int d^3y \sum_{\mu,\nu}F_{\mu\nu}(x)
{1\over |x-y|^2} F_{\mu\nu}(y),
\label{LA}
\end{equation}
where $F_{\mu\nu}=\partial_\mu A_\nu-\partial_\nu A_\mu$.
Similarly the effective action of the field $\sigma_x$ is obtained as 
\begin{equation}
{\cal L}_\sigma \propto \int d^3p\; \tilde{\sigma}(-p){1\over |p|}
\tilde{\sigma}(p),
\label{Lsigma}
\end{equation}
where $\tilde{\sigma}(p)$ is the Fourier transformed field of $\sigma_x$.
Equation (\ref{Lsigma}) shows that fluctuations of the field $\sigma_x$
are strongly suppressed at large distances.
In the $CP^1$ model on the $3$-dimensional space-time cubic lattice, 
a similar expression of the action ${\cal L}_A$
and ${\cal L}_\sigma$ is obtained by the hopping expansion of $z^a_x\;(a=1,2)$.
In the effective action $S_A$ of the compact gauge field $U_\mu=e^{iA_\mu}$, 
the following nonlocal
terms appear 
\begin{equation}
S_A\sim \sum_\Gamma\gamma^{|\Gamma|}\prod_\Gamma U_\mu (x),
\label{SA}
\end{equation}
where the summation over closed loops $\Gamma$
includes loops of an arbitrary large size, $|\Gamma|$ is the length
of $\Gamma$ and the parameter $\gamma$ is estimated as 
$\gamma\sim{1\over 2d}$ for massless $z^a_x$
with the dimension of the lattice $d=3$.
We shall focus our interest on the gauge dynamics of the above nonlocal
action which is one of the most important problems in the theoretical studies
on the strongly-correlated electron systems in these days
and is still controversial.
At present it is known that there exists only the confinement phase in the 
$(2+1)$-dimensional compact U(1) gauge theory {\em without
matter couplings}.\cite{polyakov}
However phase structure is not clear when the U(1) gauge field couples to 
matter fields, particularly gapless matter 
fields\cite{gauge,gauge1,gauge2,gauge3,gauge4}.
In particular in Ref.\cite{gauge,gauge1}, it is argued that a deconfinement
phase is realized by the gapless fermion couplings.
Simple mean-field type argument is {\em not} applicable for the nonlocal gauge
systems and numerical studies are required.

We shall study lattice gauge model with a nonlocal action which is related
with (\ref{SA}) but slightly more tractable.
The summation over $\Gamma$ in $S_A$ (\ref{SA}) becomes 
(logarithmically) divergent for the
configulation $U_\mu=1$\cite{FN1} since
the massless relativistic
bosons $z^a_x\;(a=1,2)$, which appear at the critical point,
give divergent hopping expansion for $U_\mu=1$.
From the above discussion we shall consider the following 
$(2+1)$ dimensional lattice gauge
model with the long-range interaction in the $\hat{\tau}$ direction,
\begin{eqnarray}
S_G&=&g_1 \sum^{N_\tau-1}_{n=1}\sum_{x,\mu=1,2} {1\over n}\;
U_{\mu}(x)W_{x+\mu}(n)
         U^\dagger_{\mu}(x+n\hat{\tau})W^\dagger_x(n) \nonumber  \\
        && +g_2\sum_{pl}\prod_{pl} U,
\label{SG}
\end{eqnarray}
where $W_{x}(n)=U_{0}(x)U_{0}(x+\hat{\tau})\cdots U_{0}(x+n\hat{\tau})$,
$N_\tau$ is the system size in the $\hat{\tau}$ direction 
and $g_i\; (i=1,2)$ are
coupling constants for the time and spatial directions, respectively.
From action (\ref{SG}), it is obvious that the gauge model under study
has nonlocal coupling in the $\hat{\tau}$ direction whereas it has 
the usual local Maxwell-type correlation in the spatial directions.\cite{FN2}
Reason why we take the action (\ref{SG}) is that the $\hat{\tau}$ direction
terms logarithmically divergent for $U_\mu=1$ and also
Monte-Carlo simulations are easier for the model than those with
full-nonlocal interaction terms. 
We think that studies of the model (\ref{SG}) give important
insight for the full-nonlocal gauge system (\ref{SA}).
More comments on this point will be given after showing the
results of the Monte-Carlo simulations of the model (\ref{SG}).

%---------------------------------------------------
\begin{figure}[ht]
\begin{center}
\leavevmode
\epsfxsize=6cm
\epsffile{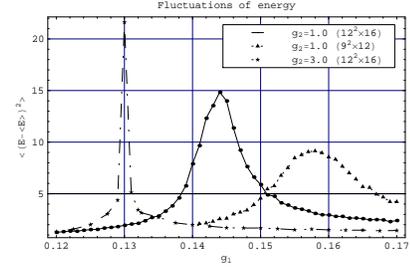}
\caption{Fluctuation of energy as a function of $g_1$ is plotted for the fixed
 $g_2=1.0$ and $3.0$. Lattice size is $12^2\times 16$ and $9^2\times 12$.
The results show the existence of a phase transition.}
\end{center}
\end{figure}
\begin{figure}[htbp]
\begin{center}
\leavevmode
\epsfxsize=4.5cm
\epsffile{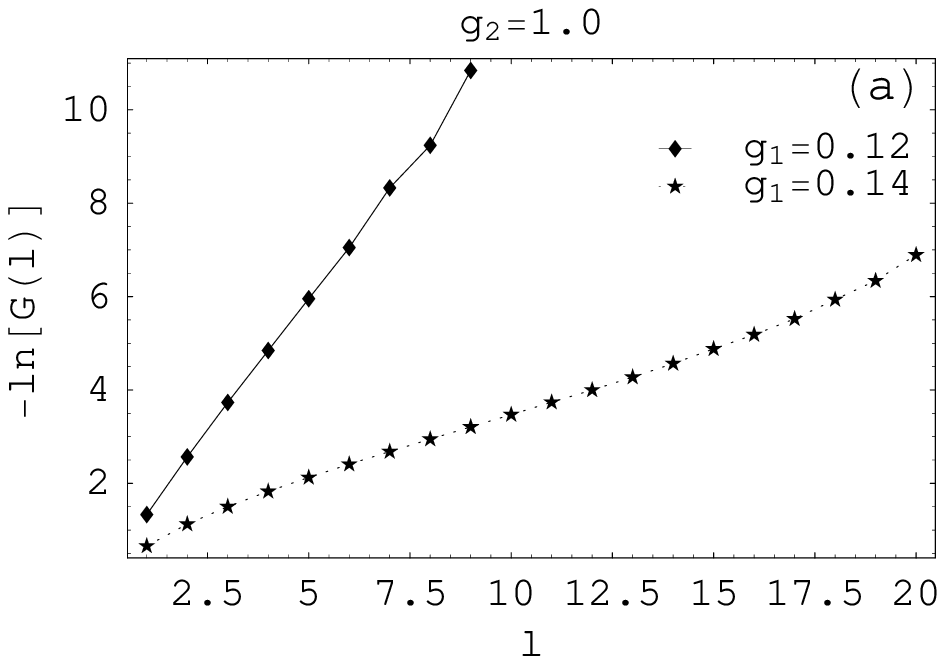}
%\end{center}
%\end{figure}
%\begin{figure}[ht]
%\begin{center}
\leavevmode
\epsfxsize=4.5cm
\epsffile{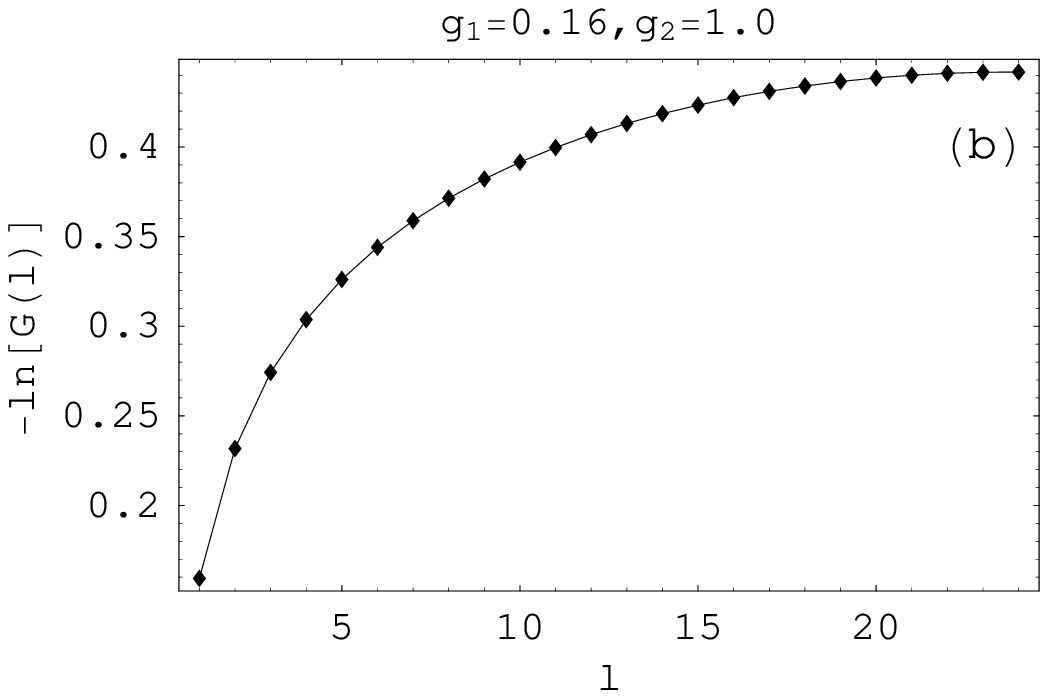}
\caption{The correlations of Polyakov lines as functions of their
 distance are plotted. The results for $g_1=0.12$ and $0.14$ are shown in
 (a), and one for $g_1=0.16$ is shown in (b). Value of $g_2$ is 
fixed as $g_2=1.0$. The results show the existence of
the confinement-deconfinement phase transition.}
\end{center}
\end{figure}
%----------------------------------------------------

We studied phase structure of the model (\ref{SG}) by the
Monte-Carlo simulations of the standard Metropolis algorithm.
We calculated the fluctuation of the energy $E$, i.e.,
$\langle (E-\langle E\rangle)^2\rangle$,
as a function of the coupling $g_1$ for fixed $g_2$.
The results are shown in Fig.2 for $g_2=1.0$ and $g_2=3.0$,
and they indicate the existence of a phase transition at critical 
coupling $g_{1C}\sim 0.14$ and $g_{1C}\sim 0.13$, respectively.
We varied the system size and verified that the peak gets sharper
as the system size increases.
This means that the transition is of second order.
In order to investigate gauge dynamics in each phase we calculated
the correlations of the Polyakov lines 
$G(\ell)=\langle P(x)P^\dagger(x+\ell\hat{1})\rangle$
where $P(x)=U_{0}(x)\cdots U_{0}(x+N_\tau{\hat{\tau}})$.\cite{FN3}
In Fig.3, we show the results of $G(\ell)$ for various values of 
$g_1$ and it is
obvious that the behavior of $G(\ell)$ changes at $g_1\sim g_{1C}$
from $\ln G(\ell) \propto \ell$ to $G(\ell)\sim \mbox{constant}$ 
for large $\ell$.
Then {\em the phase transition from the confinement to the
deconfinement phases occurs} as $g_1$ increases.

Numerical calculations show that the spatial coupling of the gauge field and 
larger system size {\em enhance} the deconfinement phase.
From the above results we expect that the full-nonlocal gauge model (\ref{SA}),
which results from the integration of the massless relativistic boson $z^a_x$
at the criticality, is in the deconfinement phase.
In fact the coupling constant $g_1 \sim 1$ corresponds to (\ref{SA})
in which the damping factor $\gamma^{|\Gamma|}$ balances the entropy
factor of the paths $\Gamma$.

In the deconfinement phase of (\ref{SG}), topologically nontrivial 
configurations are suppressed and the 
field-theory result (\ref{LA}) gives a qualitatively correct picture.
Charges interacting through $A_\mu$ have the potential 
$V(r) \propto 1/r$ where $r$ is the spatial distance 
between the two charges.

Let us comment on the effects of the Berry phase.
Since ${\cal L}_B$ is neglected in (\ref{SG}),
one may doubt the deconfinement phase transition observed above.
However, as the Berry phase generates extra phases for 
topologically nontrivial configurations in the path integral, 
{\em the Berry phase enhances the deconfinement}. 
In fact without these extra phases, all
instanton configurations contribute additively to disorder the gauge system. 
Thus the existence of the deconfinement phase in the gauge system (\ref{SG})
guarantees its existence even in the presence of the Berry phase.
Similar argument was used for the deconfinement transition 
at finite $T$\cite{susskind}, which is established at present.

We summarize the phase structure of the original spin model.
In the region $\alpha<\alpha_C
\;(f_{\mbox{\footnotesize{eff}}}<f_C)$, the low-energy excitations are the
massless spin waves whereas in the region $\alpha>\alpha_C
\;(f_{\mbox{\footnotesize{eff}}}>f_C)$, they are $s=1$ excitations $\vec{n}_x$.
On the critical point $\alpha=\alpha_C \;(f_{\mbox{\footnotesize{eff}}}=f_C)$,
the gauge dynamics is in the ``Coulomb" phase and
the low-energy excitations are the $s=1/2$ bosonic spinons $z^a_x\;(a=1,2)$ 
which are interacting with each other by the potential $1/r$.
The spin correlation function decays algebraically both in the N\'eel state
and at the criticality but exponent is different.
In the N\'eel state, the spin operator is given as $\vec{S}_x={1\over 2}
z^\dagger_x \vec{\sigma}z_x \sim \langle z^\dagger_x\rangle \vec{\sigma}z_x
\sim z^1_x$ whereas at the criticality $\vec{n}_x$, the bilinear of 
$z^a_x$ and $z^{a\dagger}_x\;(a=1,2)$.
Phase structure of the nonuniform AF Heisenberg model
is schematically shown in Fig.4.
%%%%%%%%%%%%%%%%%%%%%%%%%%%%%%%%%%%%%%%%%%%%%%%%%%%
%---------------------------------------------------
\begin{figure}[ht]
\begin{center}
\leavevmode
\epsfxsize=4cm
\epsffile{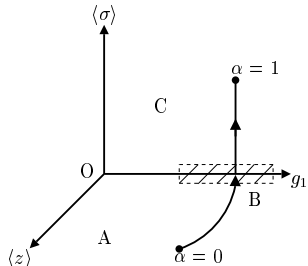}
\caption{Schematic phase diagram of the Heisenberg model. 
Phases A,B and C are the Higgs (N\'eel), Coulomb (critical) 
and confinement (dimer) phases, respectively.}
\end{center}
\end{figure}

Recently a similar phase transition from the N\'eel to the dimer
states was discussed\cite{senthil}.
There they conclude that instanton effects are irrelevant at the critical
point.
Our numerical investigation is consistent with their result but 
our study shows the
long-range interactions of the gauge field, which appear as a result of the
coupling to the massless boson $z^a_x$, play an essentially important role.
Results of more detailed studies on the long-range gauge theories
will be reported in near future.\cite{AIMS}

%%%%%%%%%%%%%%%%%%%%%%%%%%%%%%%%%%%%%%%%%%%%%%%%%%%%%%%%%%%%%%%%%%%%%%%%%%%%%%%%


\begin{references}
\bibitem{FN1}In the strong-coupling phase, the field $z_x$ acquires
nonvanishing mass by $\langle\sigma_x\rangle\neq 0$.
In this case, the summation over $\Gamma$ is convergent even for 
$U_\mu=1$.
\bibitem{FN2}The model (\ref{SG}) can be extended for the strong-coupling
phase in which $\langle\sigma_x\rangle \neq 0$.
The nonlocal terms become as 
$\sum_{n=1}{1\over n}e^{-|\langle\sigma\rangle| n}
 U_{\mu}(x)W_{x+\mu}(n)
         U^\dagger_{\mu}(x+n\hat{\tau})W^\dagger_x(n)$.
\end{references}
\end{document}